\documentclass[twocolumn,english,preprintnumbers,amsmath,amssymb,PRB,superscriptaddress,showpacs]{revtex4}

\usepackage{graphics}
\usepackage{amssymb}
\usepackage{amsmath}
\usepackage{overpic}

\usepackage{psfrag}

\newcommand{\be}{\begin{equation}}

\newcommand{\ee}{\end{equation}}

\newcommand{\bea}{\begin{eqnarray}}

\newcommand{\eea}{\end{eqnarray}}

\def\a{\alpha}

\def\d{\delta}

\def\s{\sigma}

\def\D{\Delta}

\def\nn{\nonumber}

\def\lb{\label}

\def\pref#1{(\ref{#1})}

\newcount\bozza \bozza=0

\ifnum\bozza=1

\newdimen\shift \shift=-2truecm

\def\lb#1{%
{\label{#1}\rlap{\kern\shift{$\scriptstyle#1$}}}}
\else\def\lb#1{\label{#1}} \fi

\makeatletter

\@ifundefined{textcolor}{}
{%
 \definecolor{BLACK}{gray}{0}
 \definecolor{WHITE}{gray}{1}
 \definecolor{RED}{rgb}{1,0,0}
 \definecolor{GREEN}{rgb}{0,1,0}
 \definecolor{BLUE}{rgb}{0,0,1}
 \definecolor{CYAN}{cmyk}{1,0,0,0}
 \definecolor{MAGENTA}{cmyk}{0,1,0,0}
 \definecolor{YELLOW}{cmyk}{0,0,1,0}
 }

\@ifundefined{definecolor}
 {\usepackage{color}}{}
\@ifundefined{definecolor}
 {\@ifundefined{definecolor}
 {\usepackage{color}}{}
}{}

\usepackage{dcolumn}

\usepackage{bm}

\usepackage{rotating}

\@ifundefined{definecolor}
 {\@ifundefined{definecolor}
 {\@ifundefined{definecolor}
 {\usepackage{color}}{}
}{}
}{}

\def\mS{\mathcal{S}}

\def\mR{\mathcal{R}}

\makeatother

\usepackage{babel}

\begin{document}

\title{Universal scaling of the order-parameter distribution in
  strongly disordered superconductors}

\author{G. Lemari\'e}

\affiliation{Laboratoire de Physique Th\'eorique UMR-5152, CNRS and
  Universit\'e de Toulouse, F-31062 France}

\affiliation{ISC-CNR and Department of Physics, Sapienza University of
  Rome, P.le A. Moro 2, 00185 Rome, Italy}

\author{A. Kamlapure}

\affiliation{Tata Institute of Fundamental Research, Homi Bhabha Rd., Colaba, Mumbai,
  400005, India}

\author{D. Bucheli} 

 \affiliation{ISC-CNR and Department of Physics, Sapienza University of
  Rome, P.le A. Moro 2, 00185 Rome, Italy}

\author{L. Benfatto}

 \affiliation{ISC-CNR and Department of Physics, Sapienza University of
  Rome, P.le A. Moro 2, 00185 Rome, Italy}

\author{ J. Lorenzana}

 \affiliation{ISC-CNR and Department of Physics, Sapienza University of
  Rome, P.le A. Moro 2, 00185 Rome, Italy}

\author{G.~Seibold}

\affiliation{Institut F\"ur Physik, BTU Cottbus, PBox 101344, 03013
  Cottbus,Germany}

\author{S.C. Ganguli}

\affiliation{Tata Institute of Fundamental Research, Homi Bhabha Rd., Colaba, Mumbai,
  400005, India}

\author{P. Raychaudhuri} 

\affiliation{Tata Institute of Fundamental Research, Homi Bhabha Rd., Colaba, Mumbai,
  400005, India}

\author{C. Castellani} 

 \affiliation{ISC-CNR and Department of Physics, Sapienza University of
  Rome, P.le A. Moro 2, 00185 Rome, Italy}

\date{\today}

\begin{abstract}

  We investigate theoretically and experimentally the statistical
  properties of the inhomogeneous order-parameter distribution (OPD)
  at the verge of the superconductor-insulator transition (SIT).  We
  find within two prototype fermionic and bosonic models for
  disordered superconductors that one can identify a universal
  rescaling of the OPD. By performing scanning-tunneling microscopy
  experiments in three samples of NbN with increasing disorder we show
  that such a rescaling describes also with an excellent accuracy the
  experimental data. These results can provide a breakthrough in our
  understanding of the SIT. 
\end{abstract}


\maketitle

\section{Introduction}

The interplay between disorder and superconductivity represents a
typical example of emerging complex behavior in the presence of
competing mechanisms. Indeed, while the former leads to localization
of the electrons, and to insulating-like transport, the latter
favors the formation of a macroscopic coherent electronic state able
to sustain a dissipationless current. While at moderate disorder
level the pairing mechanism persists almost unchanged\cite{anderson},
as disorder increases the superconducting (SC) critical temperature
$T_c$ decreases and ultimately a full insulating state is reached. The
most interesting case occurs when the superconductor-insulator
transition (SIT) is somehow direct, ie without an intermediate
bad-metallic state. Indeed, in this situation one can expect a
persistence of SC correlations in the insulating state and conversely
precursor effects of the insulating order on the SC
side\cite{goldman98,gantmakher}.

In the last few years considerable theoretical and experimental
advances have been made to put such a scenario on solid
grounds. In particular, new insights have been offered by experiments
of scanning tunneling
microscopy\cite{sacepe09,mondal11,sacepe11,chand12,noat12}, that have
access to the local density of states (DOS) of homogeneously strongly
disordered superconductors. The most striking features are the
emergence of an intrinsic mesoscopic inhomogeneity in the local SC
properties, and the occurrence of a large scale spectral gap $\Delta_P\gg T_c$ for
the DOS suppression, that persists well above $T_c$. These effects are
understood qualitatively by using prototype models of disordered
superconductors\cite{feigelman_review}, that can be based either on a
fermionic\cite{ma_prb85,randeria01,feigelman07,dubi07,dubi08,randeria11,seibold12}
or bosonic\cite{fisher90,ioffe_prl10,feigelman_prb10} description of
the relevant degrees of freedom.  In the former case it has been
demonstrated that a large spectral gap $\Delta_P$ survives across the
SIT, where the transition is then controlled by the presence (or absence) of global phase
coherence. Indeed, low-lying excitations living in the SC islands
that emerge in the inhomogeneous SC landscape, lead to a finite
excitation gap despite a general decrease of the SC order
parameter. Phase fluctuations made possible by the fast suppression of
the superfluid stiffness are then responsible for the SIT towards a
non-SC state with a finite $\Delta_P$. In the case of bosonic models
the focus has been put instead on a SIT driven mainly via the
localization of preformed pairs, due to quantum fluctuations
associated to the random local energies. Within this scenario, 
a glassy-like behavior of the SC state at the verge of the SIT has
been predicted\cite{ioffe_prl10,feigelman_prb10}. A typical
manifestation of such a behavior is the emergence of a
universal power-law decay of the probability distribution of the local
order-parameter values.  However, their results have been obtained by
means of a cavity approach on the Caley tree, where the number of
neighbors grows exponentially with the distance, making this lattice
structure effectively infinite dimensional. One could then wonder what
survives of the glassy-like physics in ordinary finite-dimensional
lattices with a small number of neighbors.

Despite valuable attempts\cite{randeria11,sacepe11} to establish
a link between theoretical predictions and experiments, a characteristic signature of the SIT which allows for
a convincing quantitative comparison between theory and
experiments is still missing. The present work aims at filling this gap, and
establishing at the same time a bridge between the two lines of
theoretical investigations mentioned above. More specifically, in
analogy with Refs.\ \cite{sacepe11,ioffe_prl10,feigelman_prb10}, we
shall investigate the properties of the order-parameter distribution
(OPD) with the focus on two-dimensional (2D) systems. By comparing the
results of numerical simulations of both fermionic and bosonic models
of disordered superconductors we demonstrate the emergence of
universal scaling properties of the OPD at the verge of the SIT. We
show that at strong disorder not only the typical order parameter (OP)
vanishes, but also the OPD gets logarithmically large. This suggests
a universal scaling of the OPD in the SC phase. Indeed, the OPD
obtained for different disorder levels collapse on the same curve by
introducing as scaling variable the logarithm of the OP, normalized to
its variance. More remarkably, the same scaling is in excellent
agreement with experimental data taken in three different samples of
disordered NbN films. Such a universal OPD differs from the one obtained 
in Refs.\ \cite{ioffe_prl10,feigelman_prb10} by a mapping from the bosonic model into the
directed-polymer (DP) model on the Cayley tree, an effectively
infinite-dimensional lattice. Instead, the universal OPD we find
appears to be related to the Tracy-Widom distribution, which emerges
naturally in the insulating phase by a mapping into the DP in \textit{finite}
dimension \cite{monthus_jsm12,monthus_jpa12}. In this respect our
results establish the crucial role of the lattice dimensionality both
for the properties of the OPD and for the possible relevance of the
mapping into the DP physics on the SC side of the SIT.

The paper is organized as follows. In section \ref{sec:method}, we articulate the theoretical and experimental strategies that we adopt to tackle the problem: 
first we introduce the theoretical models and how we address them numerically and then we give a description of the experimental setup. 
In section \ref{sec:theo} we discuss the numerical results for the OPD
of the theoretical models considered and their rescaling into an
universal distribution, well fitted by the Tracy-Widom
distribution. In section \ref{sec:expe} we report the experimental
data for the OPDs of the three different samples of NbN films
considered and show that they too can also be rescaled to a universal
distribution which matches very well both the numerical data and the
Tracy-Widom distribution. Our concluding remarks are reported in
Sec. \ref{sec:conclusion}.

\section{Methodology} \label{sec:method}

\subsection{Theory}

\subsubsection{Fermionic model}
 The first prototype fermionic model for a disordered
superconductor that we will analyze is the Hubbard model with random on-site energies:\cite{randeria01,seibold12}
\begin{equation}
H = -t\sum_{<ij>,\sigma}
(c_{i\sigma}^{\dag} c_{j\sigma} + h.c.) + \sum_{i,\sigma} (V_{i}-\mu )
n_{i\sigma} -|U|\sum_{i} n_{i \uparrow} n_{i \downarrow}.
\label {eq:hamil}
\end{equation}
Here $c_{i\sigma}^{\dag}$ ($c_{i\sigma}$) is the creation
(destruction) operator for an electron with spin $\sigma$ on a site
${\bf r}_i$ of a square lattice with lattice spacing $a = 1$, $t=1$ is
the nearest-neighbor hopping, $|U|$ is the pairing interaction,
$n_{i\sigma} = c_{i\sigma}^{\dag} c_{i\sigma}$, and $\mu$ is the
chemical potential.  The on-site potentials $V_{i}$ are independent
quenched random variables {which are, unless specified,} box
distributed between $-V_0$ and $V_0$, with $V_0$ denoting the disorder
amplitude.

We will investigate the model \pref{eq:hamil}  by means of 
Bogoliubov-de Gennes (BdG) mean field theory 
\cite{degennes,randeria01,seibold12}, allowing for spatial
fluctuations of {the pairing amplitude} $\Delta_i \equiv \vert U \vert
\langle c_{i \downarrow} c_{i \uparrow} \rangle$. Even though one cannot describe the SIT
within the BdG mean-field approach, it
captures already several features of strongly-disordered
superconductors\cite{randeria01,shepelyansky,dubi07}, such as the emergence
of spatial inhomogeneity of the OP and the survival of a large spectral
gap due to the interplay between superconductivity and disorder. In addition, it has recently been
shown\cite{seibold12} that at strong disorder the SC current
follows a non-trivial percolative pattern, reminiscent of the
glassy behavior suggested by the analysis of Refs.\
\cite{ioffe_prl10,feigelman_prb10}. It is then worth investigating if
also the OPD shows any particular feature at strong disorder that can
be reminiscent of the peculiar power-law decay obtained in Refs.\ 
\cite{ioffe_prl10,feigelman_prb10} when a 
non-self-averaging behavior emerges. As we shall see below, we do find
indeed a universal behavior of the OPD, which differs however
from the one obtained in Refs.\
\cite{ioffe_prl10,feigelman_prb10} on the Cayley tree.

We investigated the Hubbard model \pref{eq:hamil} in a wide
range of parameters: averaged density $\langle n \rangle \in
  [0.3,1]$, interaction strength $\vert U \vert/t \in [1,9]$, disorder
  amplitude up to $V_0/t=8$ ($t=1$ in the following), and lattices of
linear dimensions up to $L=36$, with a large number of disorder
configurations (up to $1920$). We notice that in order to investigate
the OPD one needs an average over a large number of samples of large
linear size, a task that cannot be reached with more refined
treatments beyond mean field such as Quantum Monte Carlo approach\cite{randeria11}.

\subsubsection{Bosonic model}

The propotype bosonic model for disordered superconductor has been
introduced in a seminal paper by Ma and Lee,\cite{ma_prb85} who observed that even if
single-particle states get localized by disorder 
superconductivity can survive if there are enough states localized in a
range of energy of order $\Delta$. In this situation one can show\cite{ma_prb85} that the
fermionic problem can be mapped into an effective $XY$-like spin Hamiltonian
\be
\lb{heff}
H_{I}=-\sum_{i}\xi_{i}\sigma_{i}^{z}-  \sum_{i,j} M_{ij}\left( \sigma^+_i \s^-_j+\s^-_i
  \s^+_j \right)
\ee
where $\sigma_i$ are Pauli matrices, 
$\xi_i$ are the (random) energies of the localized states  and
$M_{ij}$ are the hopping amplitude between the Cooper pairs, proportional
to the overlap between the localized states labeled $i$ and $j$ and which becomes short-ranged as disorder
increases \cite{randeria01}. In
the language of Eq.\ \pref{heff} a state with $\sigma_{i}^{z} = \pm 1$ corresponds to a site
occupied or unoccupied by a Cooper pair, while the superconducting phase corresponds to the existence
of a spontaneous magnetization in the $x-y$ plane. In the insulating phase, disorder suppresses the
pair hopping and the spins are randomly aligned along the $z$ axis. In
such a  picture the main emphasis is then placed on the competition
between local pairing and single-particle localization, and not on the role
of phase fluctuations. These are anyway allowed in the model \pref{heff}
which has the full $XY$ symmetry of the SC problem. In this respect one can
expect\cite{ioffe_prl10,feigelman_prb10}  that the main mechanism driving the SIT is also captured by a
simplified Ising version of Eq.\ \pref{heff}, 
\begin{equation}
H_{I}=-\sum_{i}\xi_{i}\sigma_{i}^{z}- g \sum_{<ij>}\sigma_{i}^{x}\sigma_{j}^{x}\label{H_I}.
\end{equation}
where we have taken $M_{ij}=g$ for nearest neighbors and $0$ otherwise. Moreover,
since the relevant quantity for the problem \pref{heff} is the ratio
$M_{ij}/\xi_i$, in the following we will take the on-site energies
$\xi_i$ as independent quenched random variables box distributed between
$-1$ and $1$, and we will control the proximity to the SIT by decreasing
$g$. 

In the present work we aim at making a quantitative comparison between the
OPDs obtained within the fermionic model \pref{eq:hamil} and the bosonic
model \pref{H_I}. In this respect, we will parametrize the results obtained
for the Hubbard model \pref{eq:hamil} in terms of an effective disorder
strength
\be
\lb{ghubb}
g=\frac{t^2}{V_0U}.
\ee
This choice is justified by the fact that in the clean case a mapping
between the model \pref{eq:hamil} and the bosonic $XY$ model
\pref{heff} can be derived at strong coupling $U \gg t$ near half filling,\cite{micnas_prb81} with an effective
hopping between Cooper pairs given by the parameter $g$ defined in Eq.\
\pref{ghubb}. Notice however that the BdG will depend in general both
on $U,V_0$ and the density: thus, the effective coupling \pref{ghubb}
must be seen as a different way to parametrize the results
obtained for fixed $U/t$ and density as a function of increasing
disorder. 
In addition, since the BdG approach neglects phase
fluctuations, a direct comparison with the
approximated Ising model \pref{H_I}, which also lacks $XY$ symmetry, is more appropriate.
In the following we will show that this approximation
is enough to describe the anomalous effects of the OPD
distribution at the verge of the SIT, since this 
physics is driven mainly by the competition between pair hopping and site
localization. This does not exclude of course that at the SIT phase fluctuations will
lead to additional remarkable effects, as discussed in
Refs.~\cite{dubi07,dubi08,randeria11,seibold12} and suggested
experimentally by measurements of the penetration depth~\cite{mondal11, Pratap13}.

The OPD for the random Ising model \pref{H_I} has been investigated in Refs.\
\cite{ioffe_prl10,feigelman_prb10,dimitrova_jsm11} by means of a  cavity mean-field
approximation on a Caley tree. This corresponds to describe the 
spin $j$ by the local
Hamiltonian
\begin{equation}
\lb{eqcav}
H_{j}^{\text{CMF}}=-\xi_{j}\sigma_{j}^{z}-\sigma_{j}^{x}\frac{g_\text{CMF}}{K}\sum_{k=1}^{K}\langle\sigma_{k}^{x}
\rangle \; ,
\end{equation}
 where $\langle\sigma_{k}^{x} \rangle$ is the magnetization at
site $k$ due to the rest of the spins \textit{in absence of $j$} and
$g_\text{CMF} \equiv K g$ is the coupling parameter considered in this
context, with $K$ the branching number of the Cayley tree.  By
defining the cavity mean field
$B_j=\frac{g_\text{CMF}}{K}\sum_{k=1}^{K}\langle\sigma_{k}^{x} \rangle$, one
gets from Eq.\ \pref{eqcav} at zero temperature that
\be
\lb{eqsj}
\langle \sigma^x_j \rangle \equiv \frac{B_j}{\sqrt{\xi^2_j+B_j^2}}
\ee
As a consequence one can write a recursion relation for the cavity field
\begin{equation}
 B_{j}=\frac{g_\text{CMF}}{K}\sum_{k=1}^{K}\frac{B_{k}}{\sqrt{B_{k}^{2}+\xi_{k}^{2}}}
 \; ,
\label{eq:mapping_Kfinite}
\end{equation}
whose solution allows one to identify a SC state as the one where
the probability distribution of the local $B_j$ admits finite values,
otherwise one recovers the insulating
state. We notice that in the CMF approach the natural quantity to
investigate is the local field $B_i$ instead of the local order parameter
$\langle \s_j^x\rangle$ given by Eq.\ \pref{eqsj}, which plays the same
role as $\Delta_i$ in the fermionic model \pref{eq:hamil}. Thus, in order
to compare the results obtained in the two approaches, we will refer in
what follows to the probability distribution of a variable $\mS_i$ which
plays in both cases the role of a normalized local field. Thus, for the CMF
one has
 \begin{equation}
\lb{defsi}
 \mS_i\equiv \frac{B_i}{g_{CMF}} \; ,
\end{equation}
while for the BdG model $\mS_i$ is given by:
\begin{equation}\label{eq:S-BdG}
 \mS_i\equiv\frac{1}{4}\sum_{k=1}^{4} \frac{2 \Delta_k}{\vert U \vert}\; ,
\end{equation}
so that $0\leq \mS_i\leq 1$ in both cases. 
Here the index $k$ runs over the $4$ nearest neighbors of site $i$.
{Notice that even though $\mS_i$ corresponds in this case to  a
coarse-graining of the pairing amplitude $\D_i$ over nearest neighbors, we
checked that there is no qualitative difference between the probability
distribution of the two quantities $\mS$ and $\Delta$ in the regime $\mS,\Delta\ll 1$ of interest. Thus, in what follows the OPD will
always refer to the probability distribution of
$\mS$.

In contrast to ordinary mean field (MF), the cavity approach allows one to
include quantum fluctuations which lead to a SIT for a finite value of the
coupling $g$.  Moreover, as we shall discuss below, a linearized version of
Eq.\ \pref{eq:mapping_Kfinite} allows for some analytical treatment of the
OPD near the SIT. The price to pay is however to work on a Cayley tree, a
cycle-free network where each node is connected to $K+1$ neighbors. The
presence of an exponentially large number of neighbors at large distance
justifies in turn the use of a mean-field approximation in Eq.\
\pref{eqcav}, that becomes exact in the large branching limit $K \gg 1$.
On the other hand, it has also been found that in 1D the predictions of the CMF approach 
coincide with the exact known results for
random Ising chains\cite{dimitrova_jsm11}. One may then wonder how sensible
the CMF results are to the lattice topology. To investigate this issue we
have considered an extension of the cavity mean-field approach introduced
in \cite{monthus_jsm12} for the 2D lattice (2D-CMF). It consists in
propagating equation \eqref{eq:mapping_Kfinite} along the diagonals only of
a 2D square lattice, so that one sums up to $K=2$ in the above cavity
equation. More specifically, one starts from a vector of boundary fields
$B_{0,y}$ for $y=1 ... L_y$ and iterates the cavity recursion
\eqref{eq:mapping_Kfinite} along direction $x$ with periodic boundary
conditions along the $y$ direction:
\begin{equation}\label{eq:2DCMF}
  B_{x+1,y}=\frac{g_\text{CMF}}{2}\sum_{k=1}^{2}\frac{B_{x,y_k}}{\sqrt{B_{x,y_k}^2+ \xi_{x,y_k}^2} } \;,
\end{equation}
where $(x,y_k)\equiv (x, y\pm 1)$ are the two preceding neighbors of $(x+1,y)$ for a
$\pi/4$ rotated square lattice, see Ref.\ \cite{monthus_jsm12}. One then
studies the boundary fields $B_{L,y}$ where $L$ is the number of
iterations of \eqref{eq:2DCMF}. It is worth noting that such
uni-directional recursion approximation neglects backward paths, even
though it includes quantum effects, so that also in this case the SIT
occurs at a finite $g_c$.  Moreover, it has been shown in
Ref.~\cite{monthus_jsm12} to describe well at least the disordered
phase $g <g_c$: here indeed the problem can be mapped into the localized phase
of the DP model, where 
directed forward paths emerge naturally, see discussion at the end of
Sec. III below. 

Finally, to make  a more direct comparison with the MF BdG approach to the
fermionic model \pref{eq:hamil}, we also solved the Ising model \pref{H_I}
  on the 2D square lattice
  by means of standard inhomogeneous  MF approach. By using the relation
  \pref{eqsj} between the magnetization and the local field we can then
  write  the self-consistent equation
\begin{equation}
 B_{j}=\frac{g_\text{MF}}{4}\sum_{k=1}^{4}\frac{B_{k}}{\sqrt{B_{k}^{2}+\xi_{k}^{2}}}
 \; ,
\label{eq:IsingMF}
\end{equation}
where $g_\text{MF} \equiv 4 g$. {Notice that in contrast to the cavity
equation \pref{eq:mapping_Kfinite} the $B_k$ and $B_j$ fields above are not indepependent,
so Eq.\ \pref{eq:IsingMF} cannot be solved recursively. The normalized
field has always the definition \pref{defsi} with $g_{CMF}$ replaced
by $g_{MF}$}. 

The bosonic model \eqref{H_I} has been investigated in a wide range of parameters: On the Cayley tree, we have considered
different branching numbers from
$K=2$ to $4$, different depths from $L=10$ to $15$ (we have verified that the distributions obtained
in the superconducting regime did not depend on $L$ \footnote{To remove the effects of surface in the CMF approach 
in the superconducting phase, 
we looked for a $L$-step self-consistent solution of the recursion relation \eqref{eq:mapping_Kfinite} where the fields at the edges all have the same distribution as the field at the root of the 
Cayley tree of depth $L$. We then verified that the resulting distribution did not depend on $L$.}),
 and disorder configurations up to $10^4$. In the 2D-CMF approach, we have considered transverse size and 
number of iterations as large as $L=L_y=5000$ (we have checked that the distributions observed in the superconducting regime
were stationary) and disorder configurations up to $10^3$, while in the 2D-MF approach we have considered lattices of linear size up to 
$L=120$ and observed no finite-size effects on the OP distributions.

\subsection{Experiments}
\begin{figure}
\includegraphics[width=\linewidth]{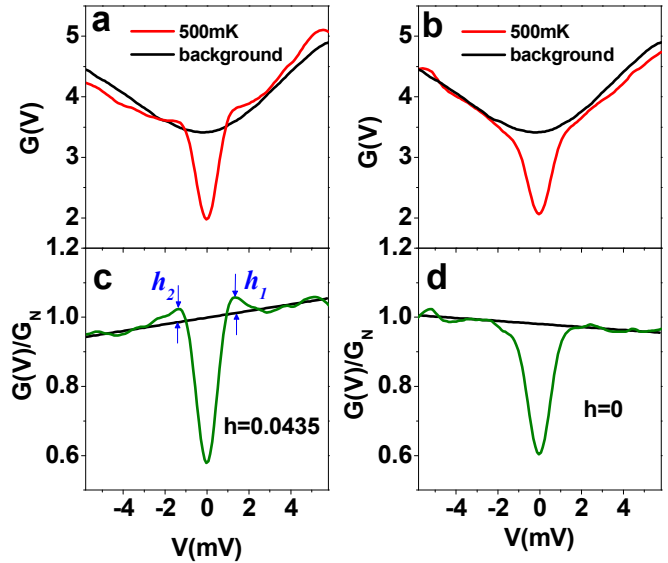}
\caption{(color online) (a)-(b) Representative spectra at 500 mK (red) at two points on the sample with $T_c \sim$ 1.65K; 
The black line shows the spatially averaged spectrum recorded at 8 K. (c)-(d) Background-corrected spectra corresponding to (a) and (b) respectively.
$h$ is the average of the coherence peak heights at positive ($h_1$) and negative bias ($h_2$), calculated with respect to the background slope determined from the conductance and high bias (black line).}
\lb{fig:exp1}
\end{figure} 

To compare the theoretical results with experiments, scanning tunneling
spectroscopy (STS) measurements were performed at 500~mK on a set of three
epitaxial NbN films grown on (100) oriented MgO substrates with different
levels of disorder. NbN is an ideal system to investigate the SIT since
disorder monotonically reduces $T_c$ \cite{mondal11} eventually giving rise
to a non-superconducting state characterized by strong superconducting
correlations \cite{chand12}. To avoid any surface contamination under
exposure to air, these samples were grown in-situ in a chamber connected to
the scanning tunneling microscope (STM). $T_c$ of the samples were measured
from resistance vs. temperature measurement after completing the STS
measurements. The samples investigated here had $T_c \sim$1.65~K, 2.9~K and
6.4~K (defined as the temperature where dc resistance goes below our
measurable limit) corresponding to an estimated $k_F l \sim$1.5, 1.8 and
2.7 respectively \cite{mondal11,chand12}.  The thickness of all films was
$\sim$50~nm which is much larger than the dirty limit coherence length
\cite{mondal_jsnm} of these films. Details of sample deposition and
characterization have been reported in Refs.\
\cite{pratap_08,chand_09,chand12}.

\begin{figure*}
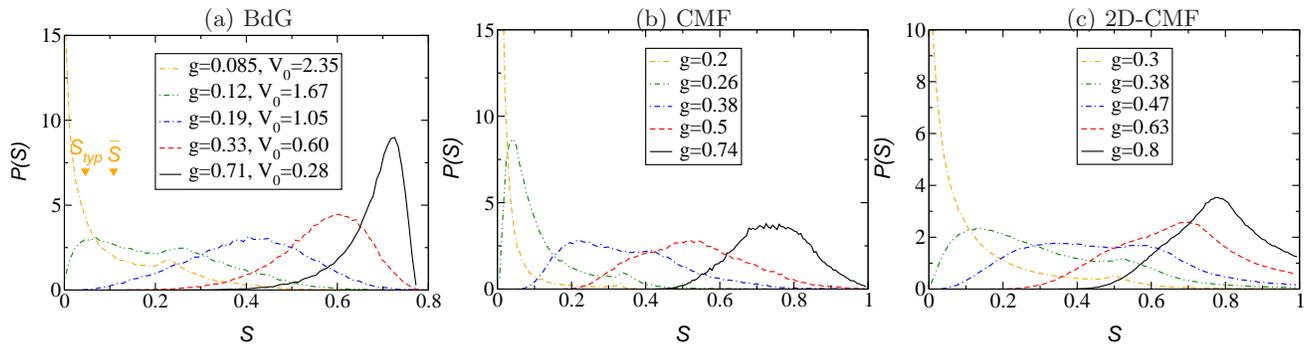

\includegraphics[width=0.32\linewidth]{Figure2a}
\put(-90,120){(a) BdG}
\includegraphics[width=0.32\linewidth]{Figure2b}
\put(-90,120){(b) CMF}
\includegraphics[width=0.32\linewidth]{Figure2c}
\put(-90,120){(c) 2D-CMF}
\caption{(color online) Distribution of the local order parameter (a) of the Hubbard model \eqref{eq:hamil} in BdG approach and of the bosonic model
\eqref{H_I} in (b) CMF and (c) 2D-CMF approaches
for different couplings $g$. 
The stronger the disorder amplitude (ie the smaller $g$), the broader the distribution
$P(\mS)$ and at a very
strong disorder, $P(\mS)$ has considerable weight near $\mS \approx 0$. The parameters are the following (see text): 
 (a) $\vert  U \vert=5$, $\langle n \rangle=0.875$, $L=25$, (b) $K=3$ and $L=10$ and (c) $L=1000$.}
\label{fig:prelim-2DBdG} 
\end{figure*}

STS measurements were performed using a home built scanning tunneling
microscope operating down to 500~mK. The construction of the STM is similar
to the one reported in ref.~\cite{mondal11}, but is based on a $^3$He cryostat
which allows us to go to lower temperatures. For each film tunneling
conductance ($dI/dV$ vs.~$V$) was measured on 32$\times$32 grid over an
area of 200$\times$200~nm. Two representative spectra on the sample with
$T_c\sim$1.65~K are shown in Figure 1 (a)-(b) (red lines). All the spectra
show a prominent dip associated with the superconducting energy gap which
adds to a broad temperature independent V-shaped background which
extends up to high bias, arising from Altshuler-Aronov type
electron-electron interactions \cite{mondal11,chand12}. At 8 K where
superconducting correlations are destroyed the spectra only show the
V-shaped background, which is spatially uniform within the noise level of
our measurements (black lines in Fig.~\ref{fig:exp1}(a)-(b)). To isolate
the feature associated with superconductivity from the background
we divide the individual spectra obtained at low temperatures by the spatially
average spectra obtained at 8K. The normalized spectra obtained in this way
(Fig.~\ref{fig:exp1}(c)-(d)) do not show a significant variation in the
magnitude of the superconducting energy gaps, but they show a large
variation in the height of the coherence peaks. Since after correcting for
background in most cases there is a small slope in the resulting spectrum, 
we fit a straight line passing through the high-bias region of the data to
get the coherence peak height (Fig.~\ref{fig:exp1}(c)-(d)), and measure the peak
heights with respect to this line for positive ($h_1$) and negative ($h_2$)
bias. We define the average height of the coherence peaks at positive and
negative bias over the normal state conductance, $h=\frac{h_1+h_2}{2}$, as
a measure of the local order parameter $\Delta_i$ for our
system\cite{sacepe11,randeria11}. To make a quantitative comparison with
the theoretical results we define for each sample the normalized local
order parameter as
 \begin{equation}
\lb{sexp}
 \mS_i^{\text{exp}}\equiv \frac{h_i}{\text{Max}[h]} \; .
\end{equation}
Thus, in analogy with the definitions \pref{defsi}-\pref{eq:S-BdG}, the
quantity $\mS_i$ is always a real number between $0$ and $1$.

\section{Numerical results}\label{sec:theo}

Let us start our analysis of the OPD with the results for the fermionic
model \pref{eq:hamil}. 
In Fig.\ \ref{fig:prelim-2DBdG} (a) we show the evolution of
the OPD $P(\mS)$ with the disorder amplitude $V_0$.  In agreement with previous
work\cite{randeria01}, the generic behavior we observe, valid for all the
parameters we considered, is an important broadening of the distribution
which gets ultimately considerable weight near $\mS \approx 0$.  As a
consequence, the typical order parameter 
\be
\lb{styp}
\mS_\text{typ}=\exp \overline{\ln
  \mS}
\ee
and the average one $\overline{\mS}$, both marked by arrows in Fig.\ \ref{fig:prelim-2DBdG} (a), 
 are very different, with $\mS_\text{typ} \ll
\overline{\mS}$. This means that the averaged quantity is governed by rare
events  and it is not representative of the typical behavior of the
system. A similar behavior is found for the bosonic model studied by either
CMF or 2D-CMF (see Fig.~\ref{fig:prelim-2DBdG} (b) and (c)).

Note the presence of a cutoff at $\mS \approx 1/4$ in the strongest
disordered case: while at $\mS\approx 1/4$ we observe a density bump, for $\mS>1/4$ the distribution falls down exponentially fast.
This is due to the small
             probability of finding
              more than one neighboring site with large pairing amplitude at strong disorder. Therefore the order parameter $\mS$, which is
defined as an average of the 4 neighboring pairing amplitudes
\eqref{eq:S-BdG}, can hardly have a value larger than $1/4$ (see also the discussion in Appendix after Eq.~\eqref{eq:ProbBlarge})}. 
The same line of reasoning applies to the 
CMF and 2D-CMF with 4 replaced by $K$, thus a cutoff at $\mS \approx 1/K$.

\begin{figure*}

\includegraphics[width=0.333333 \linewidth]{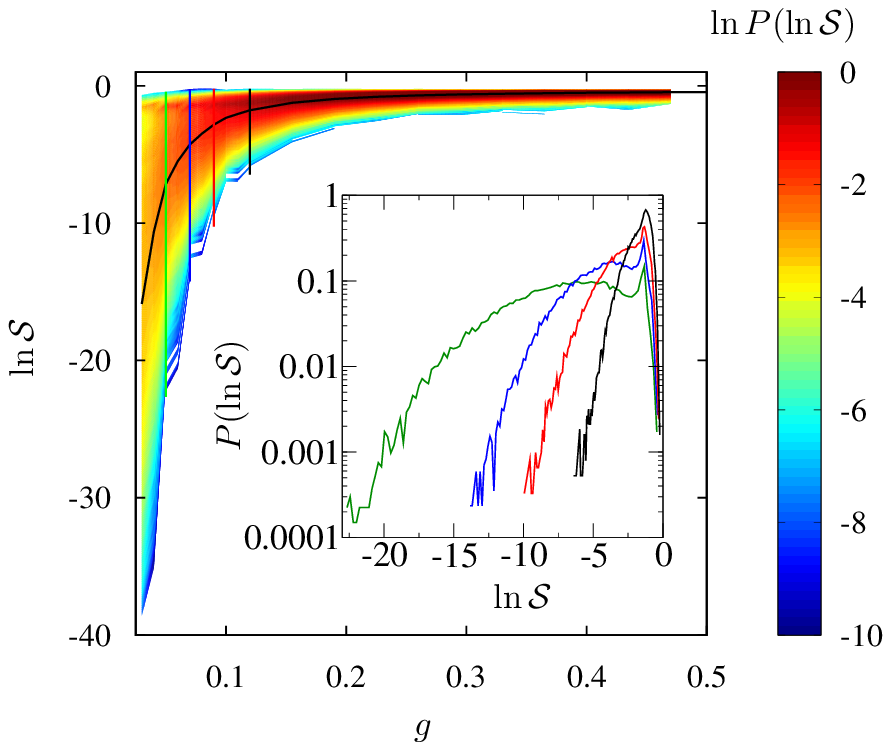}
\put(-90,140){(a) BdG}
\includegraphics[width=0.333333 \linewidth]{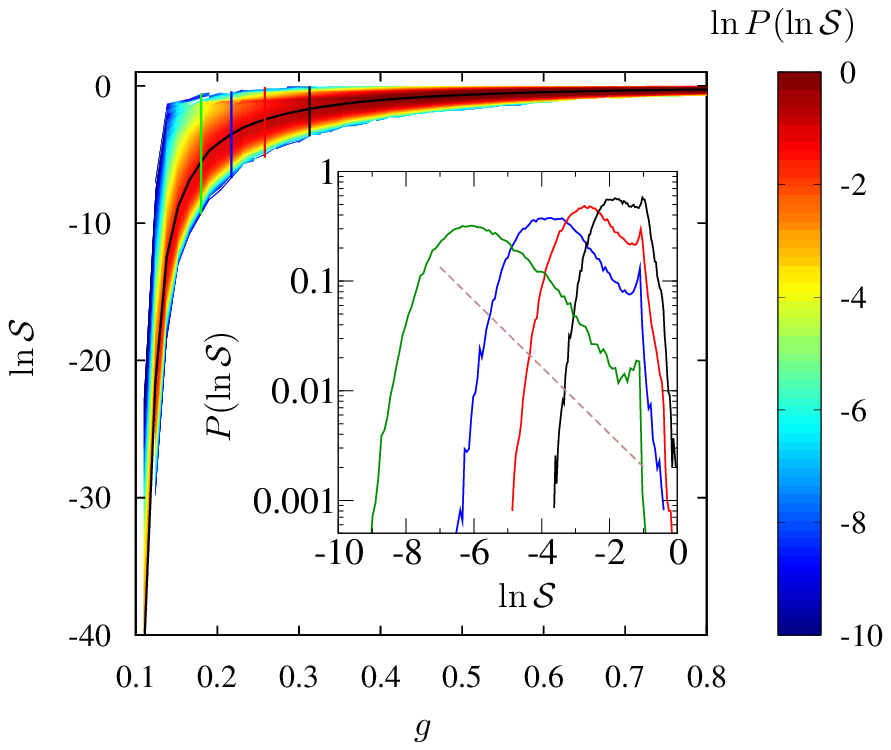}
\put(-90,140){(b) CMF}
\includegraphics[width=0.333333 \linewidth]{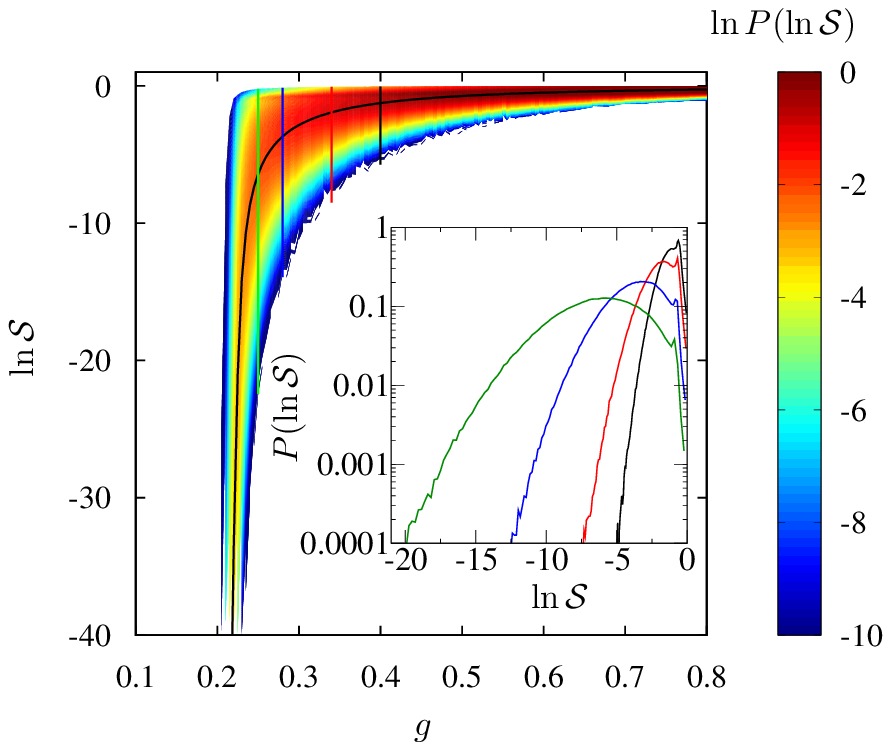}
\put(-90,140){(c) 2D-CMF}
\caption{(color online) Disorder-dependence of the OPD within (a) BdG, (b) CMF and (c) 2D-CMF.
The probability for each $\mS$ scales as the
  color code shown on the right of each panel. The maximum of the OPD is located approximately at
the typical OP, $S_{typ}$, whose $g$ dependence is shown with a
continuous line in the main panels. The insets show explicitly the $\ln \mS$ dependence
of the OPD for selected representative $g$ values in the superconducting
phase, marked by vertical bars in the main panels. Notice that in CMF a
power-law behavior $P(\ln \mS) \sim \mS^{-0.7}$ sets 
in for large OP values, $\mS_\text{typ} \ll \mS \ll g/K $ as predicted in \cite{feigelman_prb10} (see the brown dashed line in the inset).
Instead within BdG and 2D-CMF one observes the formation of
strongly asymmetric distributions with large tails extending towards
small $\mS$ values. The parameters are the following (see text): 
(a) $\vert  U \vert=5$, $\langle n \rangle=0.875$, (a) $L=10$ and (c) $L=1000$.}
\label{fig:2D} 
\end{figure*}

\begin{figure}[!b]
\begin{center}
\includegraphics[width=\linewidth]{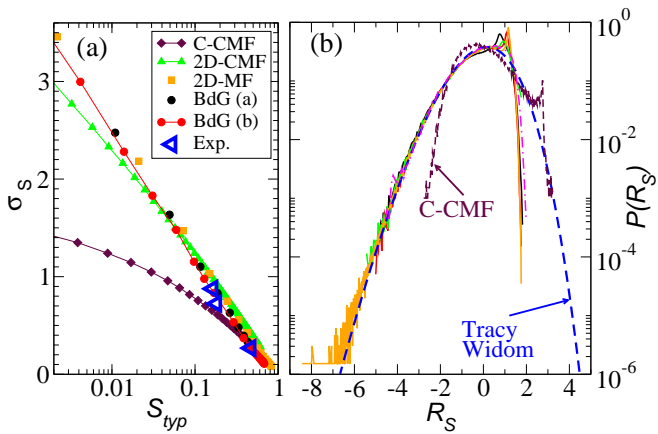}
\end{center}
\caption{(color online) (a) Evolution of the typical OP $\mS_{typ}$ and of the
  distribution width $\s_\mS$ with increasing disorder
  (ie decreasing $S_{typ}$) for 2D-CMF, 2D-MF, BdG and CMF. Notice
  that while within CMF $\sigma_{\mS}$ saturates for increasing
  disorder the 2D results show all an increasing $\s_\mS$. We also
  report the three points corresponding to the NbN samples analyzed in
  Fig.\ \ref{fig:exp3} below. (b) Rescaling of the OPD with respect to
  $S_{typ}$ and $\s_S$. All the 2D results collapse into one single curve,
  well fitted by the Tracy-Widom distribution with opposite asymmetry (see text), while the CMF
  results follow a different behavior. The
  parameters are the following: (a) 2D-CMF, $L=1000$;
  2D-MF, $L=120$; CMF, $L=15$, $K=3$; BdG (a), $\vert U \vert=9$, $\langle n
  \rangle=0.3$, $L=25$; BdG (b), $\vert U \vert=5$, $\langle
  n \rangle=0.875$, $L=25$. (b) same parameters as in (a) with in addition, 2D-CMF, $g=0.4$;
  2D-MF, $g=0.2$; CMF, $g=0.2$; BdG (a), $g=0.1$, ie $V_0=1.1$; BdG (b), $g=0.08$, ie $V_0=2.5$; magenta dashed dotted line, BdG with 
$\vert U \vert=1.5$, $\langle n
  \rangle=0.875$, $L=25$, $g=0.2$, ie $V_0=3.33$.}
\label{fig:rescaled} 
\end{figure}

In Fig.\ \ref{fig:2D} we show all our numerical results for the
fermionic and bosonic models as a function of the coupling parameter
$g$. We plot $P(\ln \mS)$ to emphasize the structure of the OPD at low
field value. The probability at each $\ln \mS$
value for the given disorder strength is represented in a color plot,
where the maximum of the distribution is located approximately at the
typical value of the OP, $\mS_\text{typ}\equiv \exp(\overline{\ln \mS})$. In
2D-CMF and CMF the SIT transition occurs at $g_c\approx 0.22$ and $g_c \approx 0.11$, 
respectively, 
while in BdG the system remains superconducting right up to $g_c=0$. 
As one can see in the insets of Fig.\
\ref{fig:2D} (a) where $P(\ln \mS)$ is reported for some representative $g$
values, for CMF we recover the expected power-law decay $P(\ln \mS)\sim \mS^{-m}$ with the 
universal (disorder-independent) exponent $m=1-eg_c\approx 0.7$ predicted in Refs.\
\cite{ioffe_prl10,feigelman_prb10}. However, such a power-law behavior is
absent in the BdG and 2D-CMF results, where 
instead $P(\ln \mS)$ appears to be dominated by the
low-field values, and to be strongly disorder-dependent. 
We notice that such a discrepancy between CMF and 2D
results can hardly be attributed to the method itself: indeed, as we
discuss in Appendix A, in 1D CMF and BdG give both
$P( \mS)\sim \mS^\alpha$ with $\alpha \rightarrow -1^+$ when $g\rightarrow g_c$ 
 which is in agreement with the
exact critical behavior of the Ising model \eqref{H_I}
\cite{dimitrova_jsm11,monthus_jpa12}. 

\begin{figure*}[!t]

\includegraphics[width=14cm,clip=true]{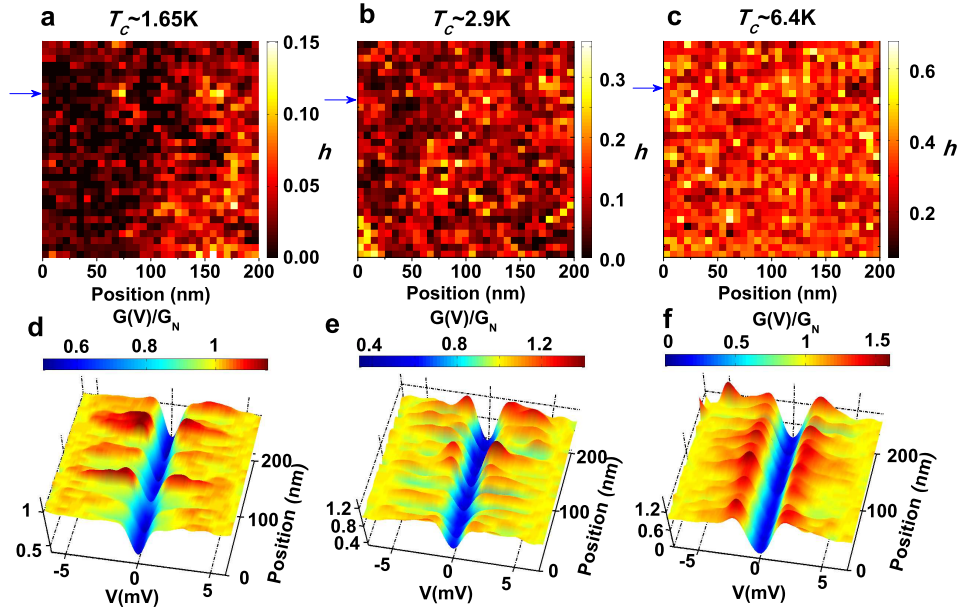}
\caption{(color online) (a)-(c) Spatial variation of the average order parameter, $h$, measured at 500 mK,  
for three NbN films with different Tc. (d)-(e) Normalized tunneling spectra
for the same samples along the line marked by arrows in panel (a)-(c).
The tunneling conductance shows the smooth variation in the height of
coherence peaks.  The linear slope from each spectrum has been 
corrected for clarity. }
\label{fig:exp2} 
\end{figure*}

Such a distinction between CMF from one side and 2D results from the
other can be made more quantitative by introducing as a scaling variable
the logarithm of the OP, normalized to its variance $\sigma_\mS^2=
{\overline{\ln^2 \mS} - \overline{\ln \mS}^2}$. Indeed, as one can see in
Fig.\ \ref{fig:rescaled}a, when disorder increases $\mS_{typ}$ and $\s_\mS$
scale in the same way in the 2D case, while within CMF $\sigma_S$ tends
to saturate at strong disorder. This result hints to a remarkable property
of the OPD, that becomes evident when the above data are rescaled with the
variable 
\be
\lb{eq:scaling2DCMF}
\mR_\mS = (\ln \mS - \ln \mS_\text{typ})/\sigma_\mS.
\ee
As shown in
Fig.\ \ref{fig:rescaled}b, provided that the coupling is small enough but
still in the SC phase, all the data (except the ones for the CMF)
collapse into one single curve (indeed, the left part of the rescaled distributions
are hardly distinguishable). We verified that such scaling holds for $\mS$ smaller than the cutoff and in
a wide range of parameters: $U$ as low as $U=1.5$, different averaged
densities, disorder strength and disorder distributions (box and gaussian
distributions). It can also be noted that a value as low as $U = 1.5$ is clearly
not in the strong coupling regime $U \gg 1$ where the fermionic Hubbard
model \eqref{eq:hamil} considered reduces essentially to hard-core
bosons. Therefore, the universal collapse onto the bosonic result validates
the bosonic scenario of Cooper pairing surviving the SIT. Another remark is
that in all the cases represented, we have chosen the longitudinal size of
the system $L$ sufficiently large to have converged to a size-independent
distribution.

\begin{figure}

\includegraphics[width=\linewidth,clip=true]{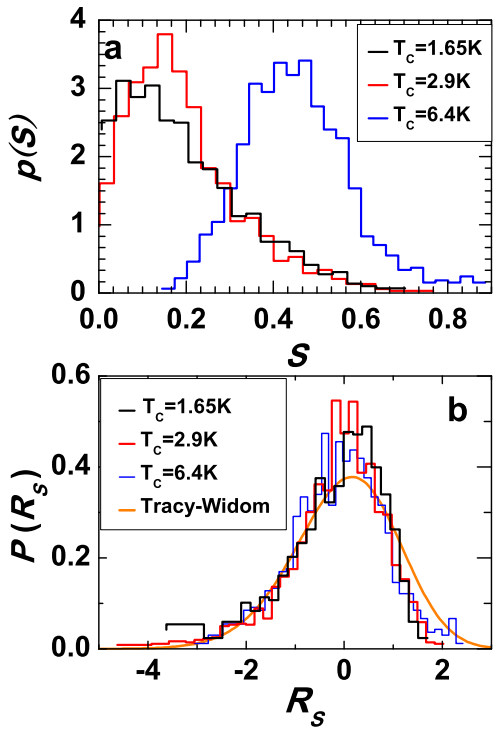}
\caption{(color online) (a) OPD of the three samples in linear scale. (b) The same data as in panel (a) plotted in terms of the rescales variable $R_\mS$. The solid line corresponds to the Tracy-Widom distribution. }
\label{fig:exp3} 
\end{figure}

As it is shown in Fig.\ \ref{fig:rescaled}b the universal distribution
followed by the rescaled data is well approximated by
the Tracy-Widom distribution\cite{spohn} with opposite asymmetry 
(ie TW$(-R_\mS)$ where TW denotes the rescaled Tracy-Widom distribution). The relevance of the Tracy-Widom
distribution in the insulating side of the SIT has been recently discussed in Refs.\
\cite{monthus_jsm12,monthus_jpa12} in connection with the physics of
DP in finite dimensions.  The connection between CMF on the
Cayley tree and the DP
physics had been already noticed in Refs.\
\cite{ioffe_prl10,feigelman_prb10}. It relies on a linearized form of the
recursion equation \pref{eq:mapping_Kfinite}, that one can expect to be
approximately valid near the SIT where $B_i$ is small:
\begin{equation}\label{eq:linearCMF}
  B_{j}\simeq \frac{g}{K}\sum_{k=1}^{K}\frac{B_{k}}{\vert \xi_{k} \vert} \;.
\end{equation}
If one then studies the value $B_0$ at the root of the Cayley tree in
response to infinitesimal fields $B_i=B\ll 1$ at the boundary, one can write
down $B_0/B$ as the sum over all the paths going from the root to the
boundary. One then sees that $B_0/B$ is exactly the partition function of a
directed polymer on a tree with edge energies $\ln \frac{ K \vert \xi_i \vert }{g}$ at temperature
$1$. Such an analogy allows one to
infer \cite{ioffe_prl10,feigelman_prb10} that at $T=0$ the
system is always in the so-called localized phase of the DP problem, or equivalently in a phase with
spontaneously broken replica symmetry (according to the language of
Refs.\ \cite{ioffe_prl10,feigelman_prb10}) where only a
small number of paths contribute to the partition function. In finite
dimension the equivalence between the cavity approximation and the DP is
only approximate: nonetheless, as it has been discussed recently in Refs.\
\cite{monthus_jsm12,monthus_jpa12}, in the insulating phase the 2D-CMF approach described above is
  strictly connected to the DP problem. More specifically, in this
  regime the OP vanishes exponentially
  fast with the system size $L$, but its fluctuations
  are tightly connected to the DP physics. Indeed, one finds that for
  a fixed scale $L$:
\begin{equation} \label{eq:DP}
 \ln \mS \approx \ln \mS_\text{typ} + \sigma_\mS \; R_\mS \;,
\end{equation}
where the variance $\sigma_\mS^2= {\overline{\ln^2 \mS} - \overline{\ln
    \mS}^2}$ scales like $\sigma_\mS \sim L^{\omega_D}$, $\omega_D$ being the
droplet exponent of the DP in $D=d+1$ dimension, 
and $R_\mS$ a random variable of order $1$ following the GOE Tracy-Widom
distribution. The droplet exponent $\omega_D$ decreases with increasing
dimensionality and it vanishes identically on the Cayley tree, which then
appears as an infinite-dimensional limit where the scaling \pref{eq:DP}
does not hold any more. When compared to our findings in Fig.\ \ref{fig:rescaled}b,
one then finds that a similar scaling holds also in the SC phase,
despite the fact that here the linearized
recursion equation \pref{eq:linearCMF} used to map into the DP model is not well justified, since 
non-linear effects due to the finite order parameter are expected to be relevant. Thus, our finding that the
OPD is related to the same Tracy-Widom distribution emerging in the DP
problem is a completely unexpected result, which shows a
posteriori that all the 2D methods and the DP problem seem to remain closely 
connected even in the SC side of the SIT transition \footnote{The appearence of TW$(-R_\mS)$ instead of TW$(R_\mS)$
is probably related to the non-linear effects which pose an upper bound
to the distribution}.

\section{Universal scaling of the experimental order parameter distribution}\label{sec:expe}

Figures \ref{fig:exp2}(a)-(c) show the spatial variation of the OP over
200$\times$200~nm area in the form of intensity plot of $h$ for the three
samples with different $T_c$. We observe a smooth variation in $h$ over
length scales of few tens of nanometers. This is further highlighted in
figures \ref{fig:exp2}(d)-(f) where we show a representative line scan of
tunneling spectra for the three samples. To make a comparison with the theoretical
results we plot in Fig.~\ref{fig:exp3}(a) the distribution of the
normalized OP $\mS^{\text{exp}}$ defined by Eq.\ \pref{sexp}. As
disorder increases one observes a steady decrease in the maximum of the OPD
along with a widening of the OPD, similar to the one reported in
Ref. \cite{sacepe11} for InO$_x$ samples. This can be further quantified by
computing $\mS^{\text{exp}}_\text{typ}$ as a function of the variance
$\sigma_\mS^\text{exp}$, which is found to follow the same trend as the
theoretical 2D results (see Fig.\ref{fig:rescaled}(a)). However, the most
striking is that by introducing the scaling variable $R_\mS$
\eqref{eq:scaling2DCMF}, all the three experimental OPD collapse into a
single universal curve, despite their apparent difference when plotted in
linear scale. In addition, the agreement with the universal Tracy-Widom
distribution found in finite dimensions is very good as well. 
We can finally note that we do not observe a sharp cutoff on the
 experimental distributions as we observed in the numerical model data.
On the other hand the experimental curves, although showing scaling
among them,  deviate from the Tracy-Widom distribution for high values of the order
parameter. This can be due to a soft cutoff effect which breaks
universality at large values of the order parameter. Alternatively it
remains the possibility that experiments do scale to an universal
curve but the Tracy-Widom distribution does  not capture the behavior
for  large $ \mS $.  More experimental and theoretical work would be
needed to clarify this issue.

\section{Conclusion}\label{sec:conclusion}

In summary we have shown both theoretically and experimentally that
the SC state at the verge of the SIT transition is characterized by a
universal behavior of the OPD. The relevant scaling variable is the logarithm of the OP normalized to
its variance. The latter
diverges as the SIT is approached, unless the problem is studied on an
infinite-dimensional lattice as the Caley tree, explaining the lack of
such universality within the CMF\cite{ioffe_prl10,feigelman_prb10}. The universal OPD shares a
pronounced similarity with the Tracy-Widom distribution, whose role in
the disordered phase of the random Ising model has been recently
discussed within the mapping into the directed-polymer model in
finite dimensions\cite{monthus_jsm12,monthus_jpa12}. Within such a
mapping additional predictions have been made, as e.g. the divergence
of the dynamical critical exponent as the SIT is approached\cite{monthus_jpa12}. This
could be tested experimentally by the critical scaling of the
superconducting fluctuations at $T_c$, as done recently in other
systems\cite{armitage}. While the critical properties of real systems
at the SIT should ultimately belong to the $XY$ universality class, at
intermediate disorder further experimental and theoretical
investigation of these predictions will further clarify the relevance
of the directed-polymer physics on the SIT.

\section{acknowledgments}
We would like to thank John Jesudasan and Vivas Bagwe for helping with the
experiments and Subash Pai from Excel Instruments, Mumbai for continuous
technical support. L.B. acknowledges partial financial support by MIUR under FIRB2012(RBFR1236VV). G.L. is supported by the EU
through a Marie Curie Fellowship, FP7-PEOPLE-2010-IEF (project number 272268).

\begin{appendix}
\section{Numerical and analytical results in 1D}
In the present appendix we will show that the
differences in the OPD between the CMF and the 2D
results presented in the paper are not due to
the underlying approximations but can be ascribed to
the different lattice structures. Indeed, in one dimension, where the Cayley-tree lattice reduces
to the usual 1D chain, both BdG and CMF approaches lead to the same
behavior of the OPD. 

\begin{figure}
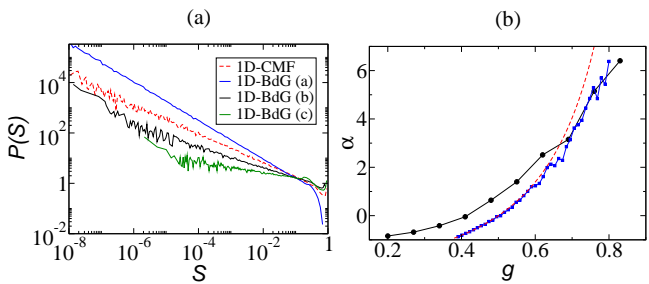

 \includegraphics[width=0.5\linewidth]{FigureA1a} 
 \includegraphics[width=0.47\linewidth]{FigureA1b}
\caption{(color online) Distribution of the superconducting order parameter in 1D. Both BdG and CMF give distributions with a universal power law shape for small values of $\mS \ll g/\sqrt{1+g^2}$. (a)  
Full lines, $P(\mS)$ given by BdG, with $g=0.37$, $L=100$ and various other parameters. From above to below:
1D-BdG (a) $\langle n \rangle=1$, $U=2$ and $V_0=1.35$; 
1D-BdG (b), $\langle n \rangle=0.7$, $U=5$ and $V_0=0.54$;
1D-BdG(c), $\langle n \rangle=0.3$, $U=9$, $V_0=0.3$. Red dashed line: $P(\mS)$ given by CMF with $g=0.436$, $L=1000$. 
(b) Power law exponent $\alpha$ of the distribution $P(\mS) \sim \mS^\alpha$
as a function of $g$. The blue squares
are CMF results, the dark bullets are BdG data with $\langle n \rangle =0.3$, $U=9$, $L=600$, 
and the dashed line is the theoretical prediction $g^{\alpha+1} (\alpha+2) = 1 $ (see text).}
\label{fig:1D} 
\end{figure}
\begin{figure}
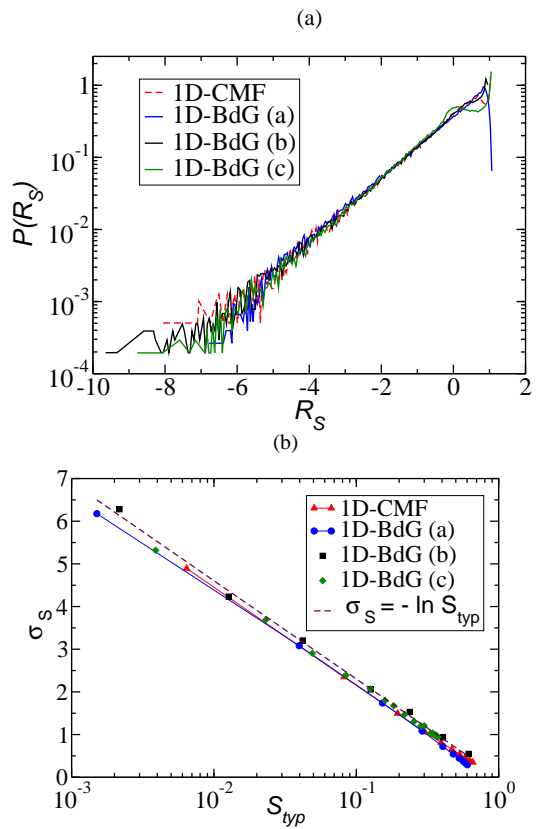

\includegraphics[width=0.8\linewidth]{FigureA2a} 
\hspace{0.5cm}
\includegraphics[width=0.75\linewidth]{FigureA2b} 
\caption{(color online) (a) Rescaling of the OPD with respect to
  $\mS_{typ}$ and $\s_\mS$. All the 1D results collapse into a single curve following the power law Eq.\ \eqref{psr}.
(b) Scaling of the distribution width vs the typical order
  parameter in 1D. The fit refers to Eq.\ \pref{defs}.
 The parameters are the following: BdG (a), $\vert U \vert=9$, $\langle n
  \rangle=0.3$; BdG (b), $\vert U \vert=9$, $\langle
  n \rangle=0.875$; BdG (c), $\vert U \vert=5$, $\langle
  n \rangle=1$.}
\lb{fig:rescaling}
\end{figure} 

In Fig.\ \ref{fig:1D} we report our results for the OPD within the
Hubbard model and CMF in 1D. As one can see, $P(\mS)$ has an universal
power-law shape for low values of $\mS <\mS_0$ with $\mS_0 $ of the
order of the effective coupling $g \equiv t^2/U V_0$: $ P(\mS)
\sim \mS ^{\alpha}$, with an exponent $\alpha$ which depends on $g$
and goes from some positive value at large $g$ (ie for small
disorder amplitude $V_0$) to $-1$ when $g \rightarrow 0$ with BdG, or
$g\rightarrow g_c=1/e$ within CMF. By universal we
mean that changing the parameters of the Hubbard model (averaged
density $\langle n \rangle$, interaction strength $U$ and disorder
strength $V_0$) does not affect the power law character but only the
exponent $\alpha$. On the contrary, for large values of $\mS>\mS_0$,
the shape of the distribution is parameter dependent. In this respect
the present 1D results demonstrate the equivalence between the BdG
approach and the CMF when the lattice structure is the same. Moreover,
such a behavior differs from the universal power-law behavior $P(\mS)\sim
\mS^{-1}$ that would be expected by extending the $K\gg 1$ result of Ref.\
\cite{ioffe_prl10,feigelman_prb10} up to $K=1$.

Once established the equivalence between the BdG and CMF results, let us
resort to the latter approach to understand analytically the power-law
evolution near the SIT. Let us start from the recursive CMF equation in 1D,
that is simply given by:
\begin{equation}
 B_{i+1}=g \frac{B_{i}}{\sqrt{B_{i}^{2}+\xi_{i}^{2}}} \; ,\label{eq:1D}
\end{equation}
From Eq.\ \pref{eq:1D} one can derive the following recursive relation for
the OPD at zero temperature ($P(\mS) = g P(B)$ since $\mS\equiv \langle \sigma^x
\rangle=B/g$ in 1D):
\begin{equation}
 P(B) = \int d B_1 P(B_1)   \int_{-1}^{1} \frac{d \xi}{2}  \delta \left( B - \frac{g B_1}{\sqrt{B_1^2 + \xi^2}} \right) \;, \label{eq:selfconsdistrib} 
\end{equation}
Since the variable $\xi$ is
independent of $B_1$ due to the recursive character of the approach, one can
integrate it explicitly. Then
one should distinguish whether $B \gtrless g^2/\sqrt{1+g^2}$. For small
values of $B\ll g^2/\sqrt{1+g^2}$, one can approximate
\eqref{eq:selfconsdistrib} by $P(B) = (g/B^2) \int_0^{B/g} d B_1 P(B_1)
B_1$.  Looking for a solution as a power law: $P(B)\sim B^\alpha$, one
obtains the condition: 
\be
\label{eqalpha}
g^{\alpha+1} (\alpha+2) = 1,
\ee
whose non-trivial solution reproduces very well the numerical data of the CMF equation in 1D
(see figure \ref{fig:1D} (b)). For large $B>g^2/\sqrt{1+g^2}$, one finds
 \begin{equation}\label{eq:ProbBlarge}
P(B) = \frac{\overline{B}/g}{(B/g)^2\sqrt{1-(B/g)^2}} \; .
\end{equation} 
Note the singularity of the distribution $P(B)$ at the cutoff $B=g$. For $K>1$ this singularity translates into a density 
bump for $P(\mS)$ at $\mS=1/K$.

As we already discussed for the 2D case in Sec.\ III, also in 1D all
the curves can be rescaled to an universal OPD by introducing the
variable $\mR_\mS = (\ln \mS - \ln \mS_\text{typ})/\sigma_\mS$, where as
before $\mS_\text{typ}\equiv \exp(\overline{\ln \mS})$ and
$\sigma_\mS^2= {\overline{\ln^2 \mS} - \overline{\ln \mS}^2}$, see Fig.\
\ref{fig:rescaling}. Also this result can be understood analytically by
assuming that the form $P(\mS)=\mS^\a (\a+1)$ for
the distribution of the order parameter holds at all $\mS$ values. Indeed
in this case it is trivial to compute
\be
\label{defs}
\overline{\ln \mS}=-\s_\mS=-\frac{1}{(\a+1)},
\ee
which demonstrates that also in 1D both $\mS_{typ}$ and $\s_\mS$ increase
as the SIT is approached. Within the same assumption we can also derive
explicitly the distribution of the variable $\mR_\mS$, given by:
\bea
P(\mR_\mS)&=&\int d\mS P(\mS) \d\left(\mR_\mS-\frac{\ln \mS-\ln
    \mS_{typ}}{\s_\mS}\right)=\nn\\
&=&(\a+1)\int d\mS \mS^\a
\frac{\delta(\mS-e^{\mR_\mS\sigma_\mS}\mS_{typ})}{1/(\mS\s_\mS)}=\nn\\
\lb{pr}
&=&\s_\mS(\a+1) \mS_{typ}^{\a+1}\exp(\mR_\mS
\s_\mS(\a+1))
\eea
On the basis of the above relation \pref{defs} we have that $\s_\mS(\a+1)=1$ and
$\mS_{typ}^{\a+1}=e^{-1}$ so that
\be
\lb{psr}
P(\mR_\mS)=\frac{1}{e}e^{\mR_\mS}
\ee

\end{appendix}


\begin{thebibliography}{cc}
\bibitem{sm}
The $^3$He refrigerator based STM used in the present work, is similar
in design to the one reported earlier in Ref.\ \cite{mondal11}.
\bibitem{notebias}
The background also displays an asymmetry between posive and
negative bias(not accounted for in Altshuler-Aronov theory) which
was also observed in NbN/insulator/Ag planar tunnel
junctions\cite{pratap09}.  This asymmetry does not appear in Ref.\
\cite{mondal11,chand12} because the background slope was removed
while aquiring the data.
\bibitem{nota_gap}
The superconducting energy gap is estimated from the position of the
coherence peaks for the first kind of spectra, and from the onset voltage
of the sharp decrease in tunneling conductance in the second kind of
spectra.
\end{thebibliography}
\end{document}